%% file: main.tex
\documentclass[runningheads]{llncs}

\usepackage[T1]{fontenc}

\usepackage{graphicx}
\usepackage{adjustbox}
\usepackage{caption}
\usepackage{subcaption}
\usepackage{float}

\usepackage[hyphens]{url}
\usepackage[hidelinks]{hyperref}
\hypersetup{breaklinks=true}
\usepackage{color}

\usepackage{booktabs}
\usepackage{bm}
\usepackage{siunitx}

\newcommand*{\permcomb}[4][0mu]{{{}^{#3}\mkern#1#2_{#4}}}
\newcommand*{\perm}[1][-3mu]{\permcomb[#1]{P}}

\begin{document}

\title{Analyzing Data Augmentation for Medical Images: A Case Study in Ultrasound Images}

\titlerunning{Analyzing Data Augmentation for Medical Images}

\author{Adam Tupper\inst{1,2} \and Christian Gagné\inst{1,2,3}}

\authorrunning{A. Tupper and C. Gagné}

\institute{Institut Intelligence et Données (IID), Université Laval \and Mila -- Quebec Artificial Intelligence Institute \and Canada-CIFAR AI Chair}

\maketitle

\begin{abstract}

Data augmentation is one of the most effective techniques to improve the generalization performance of deep neural networks. Yet, despite often facing limited data availability in medical image analysis, it is frequently underutilized. This appears to be due to a gap in our collective understanding of the efficacy of different augmentation techniques across medical imaging tasks and modalities. One domain where this is especially true is breast ultrasound images. This work addresses this issue by analyzing the effectiveness of different augmentation techniques for the classification of breast lesions in ultrasound images. We assess the generalizability of our findings across several datasets, demonstrate that certain augmentations are far more effective than others, and show that their usage leads to significant performance gains.

\end{abstract}

\section{Introduction}

Data augmentation is an essential component of deep learning. It not only improves generalization, but it is also a core component of many self- and semi-supervised learning algorithms. However, while data augmentation is ubiquitous for training deep neural networks on natural images (i.e., images of human-scale scenes captured by ordinary digital cameras), when it comes to training such models on medical images its proper usage is not as common and clearly understood \cite{chlap2021,garcea2023}. This is despite the difficulties we face collecting sufficient data, due to privacy protections, and high acquisition and annotation costs.

The underutilization of augmentation when working with medical images suggests a weaker understanding of the effectiveness of possible operations and strategies. Often, we simply apply photometric and geometric transforms proposed from natural images as is, without rigorous testing. However, the low uptake indicates that findings from natural images may not translate to medical images. This is not surprising given that the size of the objects of interest and the relevance of specific textures may differ significantly for doing detection, classification or segmentation tasks from natural images compared to other domains such as microscopic, X-ray, and ultrasound images, to name a few.

Contributing to the gap in our understanding of effective augmentation for medical images is a lack of comparative studies with controlled experiments that compare different techniques for different tasks, datasets, and modalities. While there are several excellent literature surveys on this topic \cite{chlap2021,garcea2023}, relying solely on surveys leaves us at risk of falling foul of publication bias (i.e., the file-drawer effect). In addition, these surveys highlight the difficulty in drawing conclusions on which transforms are most effective, since there are many confounding variables.

Ultimately, drawing conclusions from literature surveys alone is not enough. The problem needs to be addressed more rigorously and systematically through an experimental approach. To this end, this work presents an approach to evaluating the effectiveness of different augmentations when applied individually and when combined, using breast lesion classification in ultrasound images as a specific case study. Our investigation provides key considerations for the use of data augmentation for this task. First, the effectiveness of individual augmentations varies both when performing the same task on different datasets and when performing different tasks on the same dataset. Moreover, we observe that the common strategy of applying a fixed sequence of stochastic augmentations yields little benefit. In the best-case scenario, this approach only compensates for poor augmentation choices as opposed to yielding substantial gains. In contrast, a consistently effective strategy is to apply a random set of augmentations from a diverse set.

The remainder of this paper is organized as follows. First, we present our methodology, followed by our analyses of data augmentation for breast lesion classification in ultrasound images. Finally, we discuss the implications of our work before comparing our approach to related studies.

\section{Methodology}

When working with images, we typically apply data augmentation in one of two ways. We apply a single or hand-crafted sequence of stochastic transforms, or we apply a randomly sampled set of stochastic transforms to each image. The former is more popular for medical images, despite the latter becoming the dominant approach for natural images. To compare these approaches, we propose a sequence of studies to evaluate the effectiveness of different augmentations individually, then paired with other augmentations, and finally when combined non-deterministically. We base our evaluations on $5 \times 2$ cross-validation, using paired $t$-tests to test for statistically significant differences \cite{dietterich1998} where it is sensible to do so given the number of comparisons. This approach to comparing machine learning algorithms is widely accepted and directly measures variation due to the choice of training data, which is particularly important this setting where the available data is limited. To account for multiple hypothesis testing, we use Holm-Bonferroni correction \cite{holm1979}.

\smallskip\noindent\textbf{Individual Effectiveness}\quad We first evaluate the individual effectiveness of each augmentation to identify which one are effective and which are not. Since the strength of nearly all transformations can be reduced to the point where they have no effect on the image, we do not consider it practical to waste statistical power identifying which augmentations \textit{with a particular strength} degrade performance. This is because when their strength is reduced they should have, at worst, a negligible effect.. Therefore, we focus on testing to see if commonly used configurations \textit{are} effective using one-sided $t$-tests.

\smallskip\noindent\textbf{Paired Effectiveness}\quad To assess the benefits of applying fixed sequences of transforms, we systematically evaluate the performance of ordered pairs of augmentations, since in some cases the order in which they are applied changes the resulting output. In particular, we examine the increase in performance from applying a second transform as opposed to applying only the first. This allows use to examine the interaction and possible compound gains. We restrict our analysis to pairs of transforms due to the explosion of possible combinations as the number of sequential operations increases. Furthermore, we omit the statistical significance testing because given $N$ augmentations this requires $\perm{N}{2}$ comparisons, which when correcting for multiple hypothesis testing it is unlikely to have the statistical power to detect significant differences.

\smallskip\noindent\textbf{Random Sampling}\quad Lastly, we evaluate the performance of sampling sets of augmentations. Specifically, we test the performance of the TrivialAugment algorithm \cite{muller2021} with different sets of augmentations. Given such sets, this algorithm uniformly samples augmentations to apply to each image. Despite its simplicity, it achieves excellent performance on natural image benchmarks \cite{muller2021}.

\section{Breast Lesion Classification in Ultrasound Images}

Classifying breast lesions in ultrasound images suffers several typical problems: limited publicly available data and little consensus on data augmentation strategy to follow. To illustrate this, we performed a survey of articles citing the popular Breast Ultrasound Image (BUSI) dataset \cite{al-dhabyani2020} with the search terms ``classification'' and ``deep learning'' on the Web of Science \cite{clarivate2024}. This produced 142 articles, from which 104 were retained\footnote{38 articles were discarded as they were either reviews (15), did not perform experiments on breast ultrasound images (10), had no full text (8), did not use deep learning techniques (2), performed data synthesis as opposed to augmentation (2), or specifically excluded data augmentation (1).}.

Since the BUSI dataset consists of only 780 images, we would expect data augmentation to be useful in improving the generalization of models. However, nearly half of all studies (45 of the 104 articles) did not use any form of data augmentation. The remaining papers predominantly used a combination of geometric and photometric distortions, but many were vague in their description of the data augmentation strategy used, omitting key details such as the strength and number of sequential augmentations that were applied to each image.

We subsequently applied our methodology to the BUSI and BUS-BRA \cite{gomez-flores2023a} datasets (both of which are publicly available). In the BUSI dataset, each image is labeled according to the presence of malignant (210 images), benign (437 images), or no tumors (133 images). In contrast, the examples in the BUS-BRA dataset are labeled according to pathology, either malignant (607 images) or benign (1268 images), and by BI-RADS rating \cite{birads} (562, 463, 693, and 157 images with BI-RADS ratings of 2, 3, 4, and 5, respectively). These datasets allow us to compare our findings for similar tasks using different data (pathology) and for the different tasks on the same data (pathology vs. BI-RADS rating).

\subsection{Augmentation Operations}

We evaluated all operations used in two or more studies from our citation search of the BUSI dataset. The only exception is singular value decomposition, which was used in two papers \cite{ahmed2021,alzoubi2023}, but is not implemented in Torchvision \cite{marcel2010} nor MONAI \cite{monai2020}. The 18 augmentations and their parameters are listed in Table \ref{tab:results}, with examples of each included in the supplementary material. As for the strength of each augmentation, we use the default ranges for each stochastic transform, which are in line with those we observed in our survey of prior work. Deterministic transforms (flipping, histogram equalization, median blurring, and Gaussian noise) are applied with a probability of 0.5.

\subsection{Evaluation Protocol}

For all experiments, the images were normalized using the per-channel mean and standard deviation calculated over the entire dataset, resized so that the shortest edge measured 224\,px, and finally centrally cropped so that the final image measured $224 \times 224$\,px before applying data augmentation. The only exception to this is when using random crop, where the center crop was replaced by random cropping. We fine-tuned the weights of pre-trained ResNet-18 models from the Torchvision library. We trained each model for 100 epochs using the AdamW optimizer with a learning rate and weight decay of $1 \times 10^{-5}$ and $1 \times 10^{-2}$, respectively, and recorded the best balanced validation accuracy.  All evaluations were performed with respect to balanced accuracy due to class imbalance. In addition, the models were trained using a weighted cross-entropy loss function using weights determined by the method proposed by Cui et al.\ \cite{cui2019}. Our source code is available at \href{https://github.com/adamtupper/medical-image-augmentation}{\color{blue}{https://github.com/adamtupper/medical-image-augmentation}}. 

\section{Results}

As for the results, we first report on the individual effectiveness of different data augmentations. Second, we examine the effects of applying fixed sequences of data augmentations on the BUSI dataset. Finally, we present the effectiveness of applying augmentations using the TrivialAugment algorithm.

\input{results_table}

\subsection{Augmentations Individual Effectiveness}

\begin{figure}[t]
    \centering
    \includegraphics[width=\textwidth]{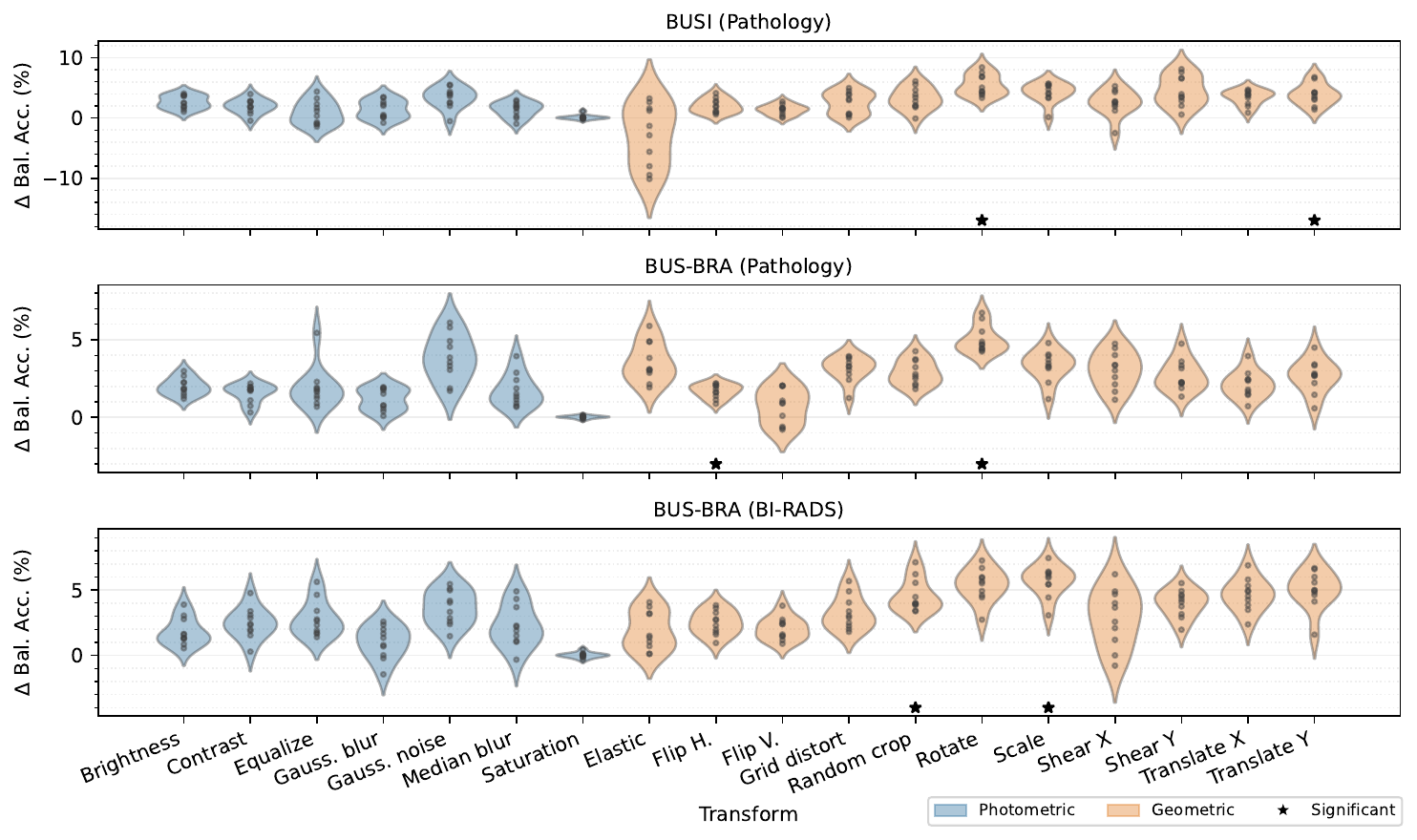}
    \caption{The change in balanced validation accuracy for each data augmentation across the three tasks.}
    \label{fig:individual_effects}
\end{figure}

Table \ref{tab:results} and Fig.\ \ref{fig:individual_effects} present the individual effectiveness of each augmentation across each dataset and task. On each of the three tasks, only two augmentations led to significantly improved balanced classification accuracy compared to no augmentation. These were rotation and Y-axis translation for the BUSI pathology task, rotation and horizontal flipping for the BUS-BRA pathology task, and scaling and random cropping for the harder BUS-BRA BI-RADS rating task. Across all three tasks, the only augmentation that decreased performance on average was the elastic transformation and only for the BUSI pathology classification task. As Fig.\ \ref{fig:individual_effects} illustrates, in all other cases we observed increases the performance, but these gains were either too small or inconsistent to be significant. Notably, none of the photometric transforms produced significant effects.

\subsection{Combining Pairs of Augmentations}

\begin{figure}[t]
    \centering
    \includegraphics[width=\textwidth]{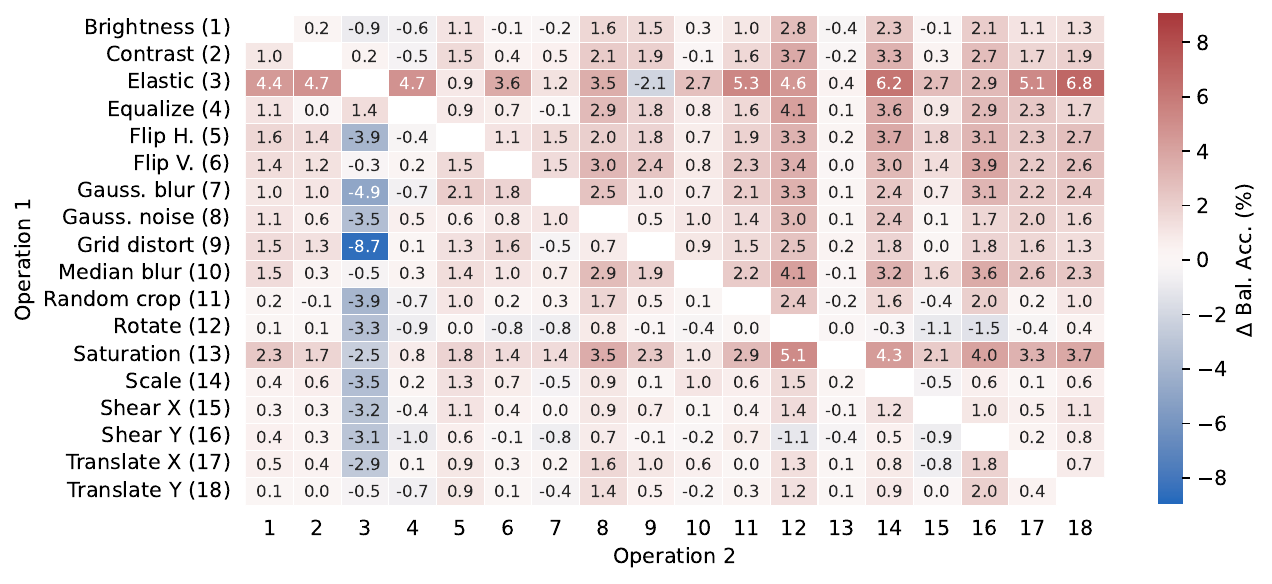}
    \caption{Change in balanced validation accuracy using two sequential operations compared to using Operation 1 alone.}
    \label{fig:paired_effects}
\end{figure}

Fig.\ \ref{fig:paired_effects} displays the relative gains and losses from apply a second random transform after the first on the BUSI pathology classification task. While in most cases the addition of a second transform led to increases in performance, the gains only tend to be substantial when applying one of the more effective augmentations after a less effective one. For example, when applying Y-axis shearing, rotation, and scaling after saturation or flipping. The only clear pattern of performance degradation occurred when elastic transformation was added, which in many cases eroded the observed gains using the first transform. This is unsurprising given the overall negative effect using elastic transform in isolation. The key result from this experiment is that while negative interactions are rare and minor, we do not observe substantial compound gains from sequentially applying individually effective augmentations.

\subsection{TrivialAugment}

\begin{figure}[t]
    \centering
    \includegraphics[width=\textwidth]{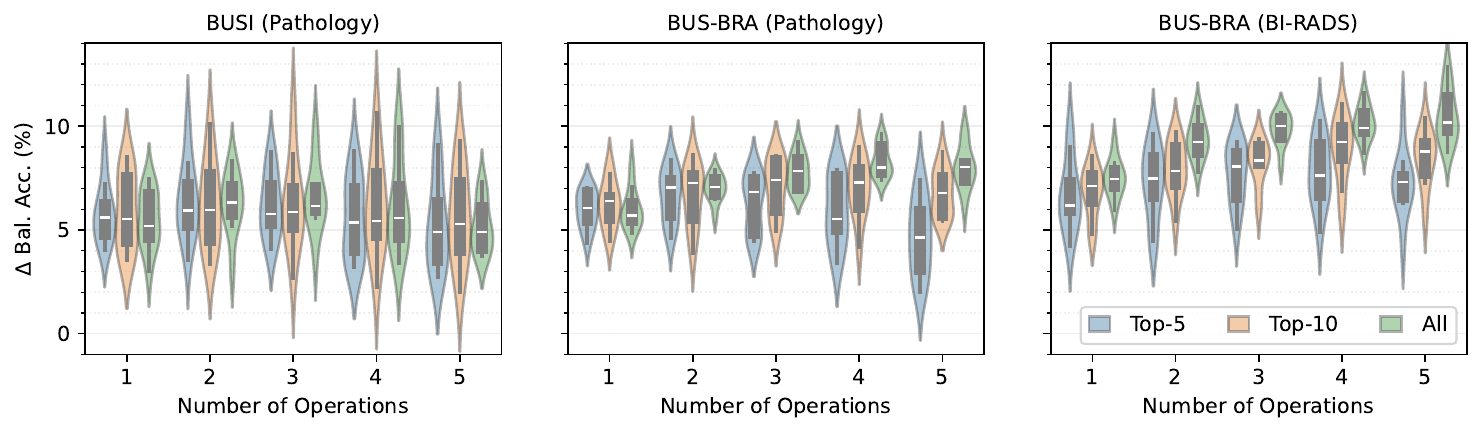}
    \caption{Change in balanced validation accuracy using TrivialAugment as the size of the augmentation pool and number of operations increased for each task.}
    \label{fig:trivial_augment}
\end{figure}

As shown in Fig.\ \ref{fig:trivial_augment}, sequentially applying a random sample of stochastic augmentations using TrivialAugment produced significant improvements in balanced accuracy compared to no data augmentation across all tasks. Exact values and significance testing results are included in the supplementary material. Despite testing TrivialAugment with the Top-5 and Top-10 augmentations for each respective task, on all three tasks the largest performance gains were using the entire set of augmentations. This highlights the benefit of sampling from a large, diverse pool of augmentations. The largest gains were observed using three ($+6.5\%$), four ($+8.3\%$), and five operations ($+10.4\%$) on the BUSI pathology, BUS-BRA pathology, and BUS-BRA BI-RADS tasks, respectively. On the BUSI task, this represents only a modest 1.1\% increase over using rotation alone, but these are substantial gains on both BUS-BRA tasks. After observing the relative performance differences between geometric and photometric transforms, we also subsequently investigated the performance of TrivialAugment using each group alone, finding that using geometric transforms alone outperformed the use of photometric transforms, but not the two combined. These results are included in the supplementary material.  

\section{Discussion}

The variation in performance across the datasets and tasks leads to several interesting takeaways:
\begin{itemize}
\item \textbf{Individual augmentation effectiveness variability}: The only individual augmentation to yield significant performance improvements in more than one task was random rotation. The differences between datasets suggests that differences in the images, for example artifacts or properties of images collected using different hardware or imaging protocols influence which augmentations are most effective, despite the modality being the same. Differences between the pathology classification task and the harder fine-grained BI-RADS rating classification task on the BUS-BRA dataset indicate that the task is also an important factor.

\item\textbf{Random augmentation selection}: Despite the observed variability in the effectiveness of each augmentation, sampling augmentations randomly from a diverse pool consistently produced the largest performance gains. Furthermore, this strategy improves as the pool of augmentations grows and with the number of augmentations applied. This leads to the finding that the individual effectiveness of each augmentation is not particularly important, a diverse pool of weak augmentations compensates for their individual relative ineffectiveness.

\item\textbf{Limitations and further research}: An element that would benefit from more rigorous analysis is the strength of augmentations. However, this adds an additional layer of complexity to the search space, limiting the type of analysis we can perform. Furthermore, our analyses are performed on only single model architecture and size, which could be generalized to different architectures and model sizes for future work. Another interesting avenue of future work is to assess augmentation effectiveness through the lens of self- and semi-supervised learning.
\end{itemize}

\section{Related Work}

Despite the clear benefits of data augmentation and its fragmented usage for medical images, there exist a limited number of comparative studies on which techniques or strategies are most effective. This has led to calls for more such studies \cite{garcea2023}. Our study extends these previous studies in several notable ways.

First, only a single previous study included ultrasound images \cite{rainio2024} and was limited to a single task. Furthermore, in this case and others \cite{bali2023,castro2018,haekal2021,hussain2018,rama2019} the effectiveness of each augmentation was only tested when used offline -- that is, used once before training to increase the size of the training set. In the case of \cite{lo2021}, augmentation policies were learned via a policy learning algorithm, but the effectiveness of individual augmentations within the defined search space was not examined. The same goes for \cite{liu2023a} who proposed an alternative augmentation strategy to TrivialAugment. Finally, \cite{eaton-rosen2018} compared the effectiveness their own sampling mixing augmentation against mixup \cite{zhang2018b}, which is frequently used in natural image settings.

It must also be mentioned there is large body of research on synthetic data generation using generative models \cite{kebaili2023}. However, this research is complementary to data augmentation as the two can be combined to increase the available training data.

\section{Conclusions}

In this work, we performed the most rigorous analysis to date of commonly used data augmentations for breast lesion classification in ultrasound images. We have demonstrated that randomly sampling data augmentations from a diverse set of or augmentations using TrivialAugment can lead to significant increases in classification performance of up to 10\% depending on the task. Overall, we hope that this work will help to establish a standard data augmentation strategy for deep learning on breast ultrasound images and provide a blueprint for future investigations into the effectiveness of other data augmentation techniques or other imaging modalities.

\begin{credits}

\subsubsection{\discintname}
The authors have no competing interests to declare that are relevant to the content of this article.
\end{credits}

\bibliographystyle{splncs04}
\bibliography{references}

\appendix
\input{supplementary}

\end{document}

%% file: results_table.tex
\begin{table}[t]
    \centering
    \begin{adjustbox}{max width=\textwidth}
    \begin{tabular}{
        l
        % BUSI
        S[table-format = 2.1] % bAcc
        S[table-format = -1.1, retain-explicit-plus] % delta bAcc
        S[table-format = <0.3,print-zero-integer=false, table-space-text-post={$^{*}$}] % P
        S[table-format = 2] % Rank
        % BUS-BRA (Pathology)
        S[table-format = 2.1] % bAcc
        S[table-format = -1.1, retain-explicit-plus] % delta bAcc
        S[table-format = <0.3,print-zero-integer=false, table-space-text-post={$^{*}$}] % P
        S[table-format = 2] % Rank
        % BUS-BRA (BIRADS)
        S[table-format = 2.1] % bAcc
        S[table-format = -1.1, retain-explicit-plus] % delta bAcc
        S[table-format = <0.3,print-zero-integer=false, table-space-text-post={$^{*}$}] % P
        S[table-format = 2] % Rank
    }\toprule
         Augmentation Strategy&  \multicolumn{12}{c}{Task}\\\cmidrule(){2-13}
         &  \multicolumn{4}{c}{BUSI (Pathology)} &   \multicolumn{4}{c}{BUS-BRA (Pathology)}& \multicolumn{4}{c}{BUS-BRA (BI-RADS)}\\\cmidrule(lr){2-5}\cmidrule(lr){6-9}\cmidrule(lr){10-13}
         & {bAcc (\%)} & {$\Delta$ bAcc} & {$P$ value} & {Rank} & {bAcc (\%)} & {$\Delta$ bAcc} & {$P$ value} & {Rank} & {bAcc (\%)} & {$\Delta$ bAcc} & {$P$ value} & {Rank}\\\midrule
         No augmentation                              & 79.0 &      &               &    & 73.3  &      &              &    & 45.8 &       &          
    &\\\cmidrule{1-13}
         \textit{Individual}\\
         Rotate ($\theta \in [-30^\circ, 30^\circ]$)  & 84.4 & +5.4 & <.001{$^{\bm{*}}$} & 1  & 78.4  & +5.1 & .003{$^{\bm{*}}$} & 1  & 51.1 & +5.3  & .023         & 2\\
         Scaling ($\times$ 0.8--1.2)                  & 83.1 & +4.1 & .014          & 3  & 76.6  & +3.3 & .150         & 4  & 51.4 & +5.6  & .001{$^{\bm{*}}$} & 1\\
         Gaussian noise ($\mu = 0, \sigma = 0.1$)     & 82.3 & +3.3 & .058          & 6  & 77.2  & +3.9 & .173         & 2  & 49.4 & +3.6  & .013         & 7\\
         Shear Y ($\theta \in [-30^\circ, 30^\circ]$) & 83.5 & +4.5 & .010          & 2  & 76.0  & +2.7 & .161         & 8  & 49.7 & +3.9  & .019         & 6\\
         Translate Y ($d \in [-0.2, 0.2]$)            & 82.9 & +3.9 & .002{$^{\bm{*}}$}  & 4  & 76.0  & +2.7 & .035         & 9  & 50.8 & +5.0  & .018         & 3\\
         Random crop                                  & 82.2 & +3.2 & .019          & 7  & 76.2  & +2.9 & .022         & 7  & 50.4 & +4.6  & .001{$^{\bm{*}}$} & 5\\
         Translate X ($d \in [-0.2, 0.2]$)            & 82.3 & +3.3 & .022          & 5  & 75.4  & +2.1 & .083         & 10 & 50.4 & +4.6  & .023         & 4\\
         Grid distortion ($m \in [-0.03, 0.03]$)      & 81.5 & +2.5 & .034          & 9  & 76.4  & +3.1 & .003         & 5  & 49.0 & +3.2  & .080         & 8\\
         Shear X ($\theta \in [-30^\circ, 30^\circ]$) & 81.5 & +2.5 & .252          & 10 & 76.4  & +3.1 & .158         & 6  & 48.6 & +2.8  & .066         & 9\\
         Contrast ($\times$ 0.5--1.5)                 & 80.9 & +1.9 & .008          & 12 & 74.9  & +1.6 & .009         & 15 & 48.2 & +2.4  & .109         & 12\\
         Brightness ($\times$ 0.5--1.5)               & 81.6 & +2.6 & .005          & 8  & 75.3  & +2.0 & .004         & 11 & 47.6 & +1.8  & .169         & 16\\
         Elastic ($\alpha=50, \sigma=5$)              & 76.1 & -2.9 & .777          & 18 & 76.9  & +3.6 & .063         & 3  & 47.7 & +1.9  & .479         & 15\\
         Flip (horizontal)                            & 81.0 & +2.0 & .012          & 11 & 75.0  & +1.7 & .003{$^{\textbf{*}}$} & 14 & 48.2 & +2.4  & .013         & 11\\
         Equalize                                     & 79.9 & +0.9 & .326          & 16 & 75.2  & +1.9 & .046         & 12 & 48.6 & +2.8  & .030         & 10\\
         Median blur ($r=3$)                          & 80.3 & +1.3 & .030          & 14 & 75.1  & +1.8 & .201         & 13 & 48.0 & +2.2  & .029         & 13\\
         Flip (vertical)                              & 80.3 & +1.3 & .023          & 15 & 74.0  & +0.7 & .224         & 17 & 47.8 & +2.0  & .005         & 14\\
         Gaussian blur ($\sigma \in [0.1, 2]$)        & 80.4 & +1.4 & .163          & 13 & 74.5  & +1.2 & .049         & 16 & 46.7 & +0.9  & .497         & 17\\
         Saturation ($\times$ 0.5--1.5)               & 79.1 & +0.1 & .431          & 17 & 73.3  & 0.0  & .961         & 18 & 45.8 &  0.0  & .757         & 18\\\bottomrule
         
    \end{tabular}
    \end{adjustbox}
    \caption{The effectiveness of individual augmentations on each task. We report the average balanced accuracy (bAcc) and change in balanced accuracy ($\Delta$ bAcc) along with the unadjusted \textit{P} values. ``*'' indicates $0.01 < P < 0.05$ after Holm-Bonferroni correction.}
    \label{tab:results}
\end{table}

%% file: supplementary.tex
\section{Illustrations of Data Augmentations}

\begin{figure}[H]
    \centering
    \includegraphics[width=\textwidth]{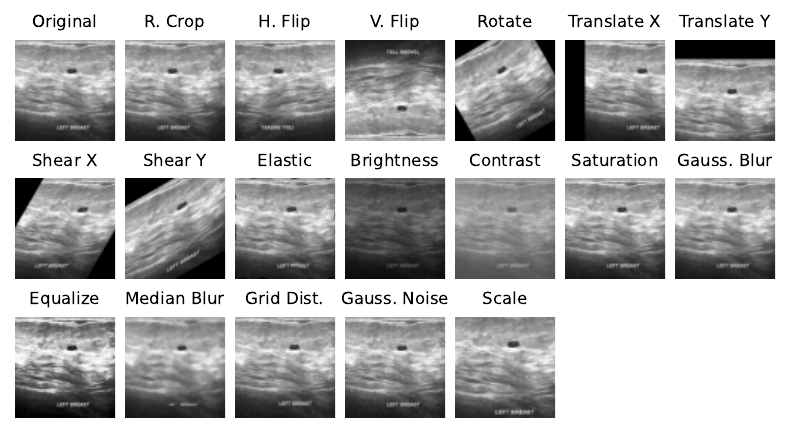}
    \caption{An illustration of the maximum strength of each augmentation when applied to an image from the Breast Ultrasound Image (BUSI) dataset.}
    \label{fig:augmentations}
\end{figure}

\section{Geometric vs. Photometric Transforms for TrivialAugment}

\begin{figure}[H]
    \centering
    \includegraphics[width=\textwidth]{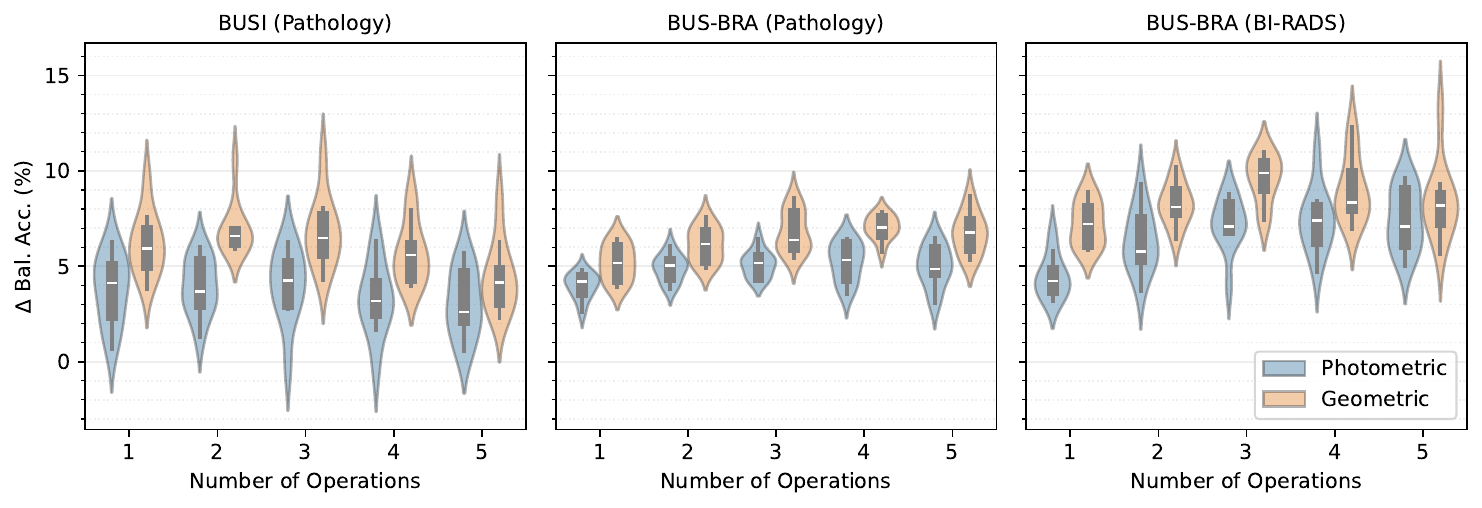}
    \caption{A comparison between the use of only photometric or geometric transforms with TrivialAugment.}
    \label{fig:augmentations}
\end{figure}

\section{Full Results using Trivial Augment}

\input{trivial_augment_table}

%% file: trivial_augment_table.tex
\begin{table}[H]
    \centering
    \begin{adjustbox}{max width=\textwidth}
    \begin{tabular}{
        l
        % BUSI
        S[table-format = 2.1] % bAcc
        S[table-format = 1.1, retain-explicit-plus] % delta bAcc
        S[table-format = <0.3,print-zero-integer=false, table-space-text-post={$^{**}$}] % P
        % BUS-BRA (Pathology)
        S[table-format = 2.1] % bAcc
        S[table-format = 1.1, retain-explicit-plus] % delta bAcc
        S[table-format = <0.3,print-zero-integer=false, table-space-text-post={$^{**}$}] % P
        % BUS-BRA (BIRADS)
        S[table-format = 2.1] % bAcc
        S[table-format = 2.1, retain-explicit-plus] % delta bAcc
        S[table-format = <0.3,print-zero-integer=false, table-space-text-post={$^{***}$}] % P
    }\toprule
         Augmentation Set  &  \multicolumn{9}{c}{Task}\\\cmidrule(){2-10}
                           & \multicolumn{3}{c}{BUSI (Pathology)}           & \multicolumn{3}{c}{BUS-BRA (Pathology)}        & \multicolumn{3}{c}{BUS-BRA (BI-RADS)}\\
                           \cmidrule(lr){2-4}\cmidrule(lr){5-7}\cmidrule(lr){8-10}
                           & {bAcc (\%)} & {$\Delta$ bAcc} & {$P$ value}    & {bAcc (\%)} & {$\Delta$ bAcc} & {$P$ value}    & {bAcc (\%)} & {$\Delta$ bAcc} & {$P$ value}\\\midrule
         No augmentation   & 79.0        &                 &                & 73.3        &                 &                & 45.8        &                 &            \\\cmidrule{1-10}
         \textit{All}\\
         1 operation       & 84.5        & +5.5            & .003{$^{*}$}   & 79.3        & +6.0            & .002{$^{*}$}   & 53.3        & +7.5            & <.001{$^{***}$}\\
         2 operations      & 85.3        & +6.3            & <.001{$^{**}$} & 80.3        & +7.0            & <.001{$^{**}$} & 55.1        & +9.3            & <.001{$^{**}$}\\
         3 operations      & 85.5        & +6.5            & <.001{$^{**}$} & 81.1        & +7.8            & <.001{$^{**}$} & 55.6        & +9.8            & <.001{$^{***}$}\\
         4 operations      & 85.1        & +6.1            & .005{$^{*}$}   & 81.6        & +8.3            & <.001{$^{**}$} & 55.9        & +10.1           & <.001{$^{***}$}\\
         5 operations      & 84.1        & +5.1            & .004{$^{*}$}   & 81.3        & +8.0            & <.001{$^{**}$} & 56.2        & +10.4           & <.001{$^{***}$}\\\cmidrule{1-10}
         \textit{Top 10}\\
         1 operation       & 84.8        & +5.8            & .006{$^{*}$}   & 79.5        & +6.2            & .003{$^{*}$}   & 52.6        & +6.8            & <.001{$^{**}$}\\
         2 operations      & 85.2        & +6.2            & .002{$^{*}$}   & 80.0        & +6.7            & .041           & 53.7        & +7.9            & <.001{$^{**}$}\\
         3 operations      & 85.2        & +6.2            & .003{$^{*}$}   & 80.4        & +7.1            & .012           & 54.0        & +8.2            & <.001{$^{**}$}\\
         4 operations      & 85.1        & +6.1            & .001{$^{*}$}   & 80.3        & +7.0            & .030           & 54.7        & +8.9            & <.001{$^{**}$}\\
         5 operations      & 84.6        & +5.6            & .004{$^{*}$}   & 80.1        & +6.8            & .007           & 54.2        & +8.4            & <.001{$^{**}$}\\\cmidrule{1-10}
         \textit{Top 5}\\
         1 operation       & 84.7        & +5.7            & <.001{$^{**}$} & 79.3        & +6.0            & .003{$^{*}$}   & 52.4        & +6.6            & .003{$^{*}$}\\
         2 operations      & 85.3        & +6.3            & <.001{$^{**}$} & 80.0        & +6.7            & .016           & 53.0        & +7.2            & .008{$^{*}$}\\
         3 operations      & 85.2        & +6.2            & .001{$^{*}$}   & 79.6        & +6.3            & .026           & 53.4        & +7.6            & .005{$^{*}$}\\
         4 operations      & 84.5        & +5.5            & .003{$^{*}$}   & 79.2        & +5.9            & .073           & 53.6        & +7.8            & .007{$^{*}$}\\
         5 operations      & 84.2        & +5.2            & .001{$^{*}$}   & 78.0        & +4.7            & .218           & 53.0        & +7.2            & .001{$^{**}$}\\\bottomrule     
    \end{tabular}
    \end{adjustbox}
    \caption{The full results of our evaluations of TrivialAugment using different sets of transforms and different numbers of operations on each task. We report the average balanced accuracy (bAcc) and change in balanced accuracy ($\Delta$ bAcc) over the five repetitions of two-fold cross validation along with unadjusted \textit{P} values for comparisons against no augmentation. ``*'', ``**'', and ``***'' indicate \textit{P} values less than 0.05, 0.01, and 0.001, respectively, after Holm-Bonferroni correction.}
    \label{tab:results}
\end{table}

%% file: main.bbl
\begin{thebibliography}{10}
\providecommand{\url}[1]{\texttt{#1}}
\providecommand{\urlprefix}{URL }
\providecommand{\doi}[1]{https://doi.org/#1}

\bibitem{birads}
Breast {Imaging} {Reporting} \& {Data} {System},
  \url{https://www.acr.org/Clinical-Resources/Reporting-and-Data-Systems/Bi-Rads}

\bibitem{ahmed2021}
Ahmed, M., AlZoubi, A., Du, H.: Improving {Generalization} of {ENAS}-{Based}
  {CNN} {Models} for {Breast} {Lesion} {Classification} from {Ultrasound}
  {Images}. vol. 12722, pp. 438--453 (2021). \doi{10.1007/978-3-030-80432-9_33}

\bibitem{al-dhabyani2020}
Al-Dhabyani, W., Gomaa, M., Khaled, H., Fahmy, A.: Dataset of breast ultrasound
  images. Data in Brief  \textbf{28},  104863 (Feb 2020).
  \doi{10.1016/j.dib.2019.104863}

\bibitem{alzoubi2023}
Alzoubi, A., Lu, F., Zhu, Y., Ying, T., Ahmed, M., Du, H.: Classification of
  breast lesions in ultrasound images using deep convolutional neural networks:
  transfer learning versus automatic architecture design. Medical and
  Biological Engineering \& Computing  (Sep 2023).
  \doi{10.1007/s11517-023-02922-y}

\bibitem{bali2023}
Bali, M., Mahara, T.: Comparison of {Affine} and {DCGAN}-based {Data}
  {Augmentation} {Techniques} for {Chest} {X}-{Ray} {Classification}. Procedia
  Computer Science  \textbf{218},  283--290 (Jan 2023).
  \doi{10.1016/j.procs.2023.01.010}

\bibitem{castro2018}
Castro, E., Cardoso, J.S., Pereira, J.C.: Elastic deformations for data
  augmentation in breast cancer mass detection. In: 2018 {IEEE} {EMBS}
  {International} {Conference} on {Biomedical} \& {Health} {Informatics}
  ({BHI}). pp. 230--234 (Mar 2018). \doi{10.1109/BHI.2018.8333411}

\bibitem{chlap2021}
Chlap, P., Min, H., Vandenberg, N., Dowling, J., Holloway, L., Haworth, A.: A
  review of medical image data augmentation techniques for deep learning
  applications. Journal of Medical Imaging and Radiation Oncology
  \textbf{65}(5),  545--563 (2021). \doi{10.1111/1754-9485.13261}

\bibitem{clarivate2024}
Clarivate: Web of {Science} (2024),
  \url{https://www.webofscience.com/wos/woscc/summary/23f8e70e-6c6e-4a90-9acd-e85a06f98a28-d0de2224/relevance/1}

\bibitem{monai2020}
Consortium, T.M.: Project {MONAI} (Dec 2020),
  \url{https://doi.org/10.5281/zenodo.4323059}

\bibitem{cui2019}
Cui, Y., Jia, M., Lin, T.Y., Song, Y., Belongie, S.: Class-{Balanced} {Loss}
  {Based} on {Effective} {Number} of {Samples}. pp. 9268--9277 (2019)

\bibitem{dietterich1998}
Dietterich, T.G.: Approximate {Statistical} {Tests} for {Comparing}
  {Supervised} {Classification} {Learning} {Algorithms}. Neural Computation
  \textbf{10}(7),  1895--1923 (Oct 1998). \doi{10.1162/089976698300017197}

\bibitem{eaton-rosen2018}
Eaton-Rosen, Z., Bragman, F., Ourselin, S., Cardoso, M.J.: Improving {Data}
  {Augmentation} for {Medical} {Image} {Segmentation}  (Apr 2018),
  \url{https://openreview.net/forum?id=rkBBChjiG}

\bibitem{garcea2023}
Garcea, F., Serra, A., Lamberti, F., Morra, L.: Data augmentation for medical
  imaging: {A} systematic literature review. Computers in Biology and Medicine
  \textbf{152},  106391 (Jan 2023). \doi{10.1016/j.compbiomed.2022.106391}

\bibitem{gomez-flores2023a}
Gómez-Flores, W., Gregorio-Calas, M.J., Coelho~de Albuquerque~Pereira, W.:
  {BUS}-{BRA}: {A} breast ultrasound dataset for assessing computer-aided
  diagnosis systems. Medical Physics pp. 1--14 (Nov 2023).
  \doi{10.1002/mp.16812}

\bibitem{haekal2021}
Haekal, M., Septiawan, R.R., Haryanto, F., Arif, I.: A comparison on the use of
  {Perlin}-noise and {Gaussian} noise based augmentation on {X}-ray
  classification of lung cancer patient. Journal of Physics: Conference Series
  \textbf{1951}(1),  012064 (Jun 2021). \doi{10.1088/1742-6596/1951/1/012064}

\bibitem{holm1979}
Holm, S.: A {Simple} {Sequentially} {Rejective} {Multiple} {Test} {Procedure}.
  Scandinavian Journal of Statistics  \textbf{6}(2),  65--70 (1979),
  \url{https://www.jstor.org/stable/4615733}

\bibitem{hussain2018}
Hussain, Z., Gimenez, F., Yi, D., Rubin, D.: Differential {Data} {Augmentation}
  {Techniques} for {Medical} {Imaging} {Classification} {Tasks}. AMIA Annual
  Symposium Proceedings  \textbf{2017},  979--984 (Apr 2018),
  \url{https://www.ncbi.nlm.nih.gov/pmc/articles/PMC5977656/}

\bibitem{kebaili2023}
Kebaili, A., Lapuyade-Lahorgue, J., Ruan, S.: Deep {Learning} {Approaches} for
  {Data} {Augmentation} in {Medical} {Imaging}: {A} {Review}. Journal of
  Imaging  \textbf{9}(4), ~81 (Apr 2023),
  \url{https://www.mdpi.com/2313-433X/9/4/81}

\bibitem{liu2023a}
Liu, Z., Lv, Q., Li, Y., Yang, Z., Shen, L.: {MedAugment}: {Universal}
  {Automatic} {Data} {Augmentation} {Plug}-in for {Medical} {Image} {Analysis}
  (Jun 2023), \url{http://arxiv.org/abs/2306.17466}, arXiv:2306.17466 [cs,
  eess]

\bibitem{lo2021}
Lo, J., Cardinell, J., Costanzo, A., Sussman, D.: Medical {Augmentation}
  ({Med}-{Aug}) for {Optimal} {Data} {Augmentation} in {Medical} {Deep}
  {Learning} {Networks}. Sensors  \textbf{21}(21), ~7018 (Jan 2021).
  \doi{10.3390/s21217018}

\bibitem{marcel2010}
Marcel, S., Rodriguez, Y.: Torchvision the machine-vision package of torch. In:
  Proceedings of the 18th {ACM} international conference on {Multimedia}. pp.
  1485--1488. {MM} '10, Association for Computing Machinery, New York, NY, USA
  (Oct 2010). \doi{10.1145/1873951.1874254}

\bibitem{muller2021}
Müller, S.G., Hutter, F.: {TrivialAugment}: {Tuning}-free {Yet}
  {State}-of-the-{Art} {Data} {Augmentation} (Aug 2021),
  \url{http://arxiv.org/abs/2103.10158}, arXiv:2103.10158 [cs]

\bibitem{rainio2024}
Rainio, O., Klén, R.: Comparison of simple augmentation transformations for a
  convolutional neural network classifying medical images. Signal, Image and
  Video Processing  (Feb 2024). \doi{10.1007/s11760-024-02998-5}

\bibitem{rama2019}
Rama, J., Nalini, C., Kumaravel, A.: Image pre-processing: enhance the
  performance of medical image classification using various data augmentation
  technique. ACCENTS Transactions on Image Processing and Computer Vision
  \textbf{5}(14) (Feb 2019). \doi{DOI:10.19101/TIPCV.2018.413001}

\bibitem{zhang2018b}
Zhang, H., Cisse, M., Dauphin, Y.N., Lopez-Paz, D.: mixup: {Beyond} empirical
  risk minimization. In: 6th {International} {Conference} on {Learning}
  {Representations}, {ICLR} 2018, {Vancouver}, {BC}, {Canada}, {April} 30 -
  {May} 3, 2018, {Conference} {Track} {Proceedings} (2018),
  \url{https://openreview.net/forum?id=r1Ddp1-Rb}

\end{thebibliography}
